\documentclass[twocolumn,showpacs]{revtex4}

\usepackage{amssymb,epsfig}
\usepackage{graphicx}
\usepackage{dcolumn}
\usepackage{bm}

\def\be{\begin{equation}} \def\ee{\end{equation}}
\def\bq{\begin{eqnarray}} \def\eq{\end{eqnarray}}
\def\p{\bullet}

\begin{document}

\title{Effects of color superconductivity on the structure and formation of compact stars}

\author{A. Drago$^a$, A. Lavagno$^b$, G. Pagliara$^a$}
\affiliation{$^a$Dipartimento di Fisica, Universit{\`a} di Ferrara and
INFN, Sezione di Ferrara, 44100 Ferrara, Italy}
\affiliation{$^b$Dipartimento di Fisica, Politecnico di Torino and INFN,
Sezione di Torino, 10129 Torino, Italy}


\begin{abstract}

We show that if color superconducting quark matter forms in hybrid or
quark stars it is possible to satisfy most of recent observational
boundaries on masses and radii of compact stellar objects. An energy
of the order of $10^{53}$ erg is released in the conversion from a
(metastable) hadronic star into a (stable) hybrid or quark star in
presence of a color superconducting phase. If the conversion occurs
immediately after the deleptonization of the proto-neutron star, the
released energy can help Supernovae to explode. If the conversion is
delayed the energy released can power a Gamma Ray Burst. A delay
between the Supernova and the subsequent Gamma Ray Burst is possible,
in agreement with the delay proposed in recent analysis of
astrophysical data.

\end{abstract}

\maketitle

The new accumulating data from X-ray satellites provide important
information on the structure and formation of compact stellar
objects. Concerning the structure, these data, 
are at first sight difficult to interpret in a unique and
self-consistent theoretical scenario, since some of the observations
are indicating rather small radii and other observations are
indicating large values for the mass of the star.
Concerning the formation scenario, crucial information are provided by
the very recent observations of Gamma-Ray Bursts (GRB), indicating the
possibility that some of the GRBs are associated with a 
Supernova (SN) explosion.
It has not yet been
clarified if the two explosions are always simultaneous or if, at
least in a few cases, a time delay can exist, with the SN preceding
the GRB \cite{Amati00,Reeves02,hjorth,rutledge,reeves2003}.

The effect of the transition to deconfined Quark Matter (QM) on
explosive processes like SNs and GRBs has been discussed by many
authors.  In particular, the possibility that deconfinement takes
place during the core-collapse of massive stars at the moment of the
bounce, has been discussed e.g. in Refs.\cite{noiJP,sannino} and this
mechanism could help the SN to explode by increasing the mechanical
energy associated with the bounce.  However, it seems more plausible
that deconfinement takes place only when the proto-neutron star (PNS)
has deleptonized and cooled down to a temperature of a few MeV
\cite{Benvenuto,Pons}. The energy released in the conversion to QM
produces a refreshed neutrino flux which can help the supernova to
explode in a neutrino-driven scheme. Finally, another scenario is
possible in which neutron stars having a small enough mass can exist
as metastable Hadronic Star (HS) if a non-vanishing surface tension is
present at the interface between Hadronic Matter (HM) and QM.  The
process of quark deconfinement can then be a powerful source for GRBs
and it can also explain the possible delay between a SN explosion and
the subsequent GRB \cite{noiapj}.

In recent years, many theoretical works have investigated the possible
formation of a diquark condensate in QM, at densities reachable in the
core of a compact star \cite{alf1,alf5,schafer}. The formation of this
condensate can deeply modify the structure of the star 
\cite{alf4,baldo1,blas1,lugo1}. We present here an extension of
the previous works, showing that it is possible to satisfy the
existing boundaries on mass and radius of a compact stellar object if
a diquark condensate forms in a Hybrid Star (HyS) or a Quark Star
(QS).  Moreover, we show that the formation of diquark condensate can
significantly increase the energy released in the conversion from a
purely HS into a more stable star containing deconfined QM.

To describe the high density Equation of State (EOS) of matter we adopt standard models
in the various density ranges.
Concerning the hadronic phase we use relativistic non-linear
models \cite{glen2,ms96}.  
At very low density we use the standard EOSs of Refs.\cite{negele,baym}.
For the QM phase we adopt a MIT-bag like model in which the formation of a
diquark condensate is taken into account.  
To connect the two phases of our EOS,
we impose Gibbs equilibrium conditions.  

It is widely accepted that the Color-Flavor Locking phase (CFL) is the 
real ground state of QCD at asymptoticly large densities.  We are
interested in the bulk properties of a compact star and we adopt the
simple scheme proposed in Refs.\cite{alf4,lugo1} where the
thermodynamic potential is given by the sum of two contributions. The
first term corresponds to a ``fictional'' state of unpaired QM
 in which all quarks have a common Fermi momentum chosen to
minimize the thermodynamic potential. The other term is the binding
energy $\Delta$ of the diquark condensate expanded up to order
$(\Delta/\mu)^2$. In Ref.\cite{alf4} the gap is assumed to be
independent on the chemical potential $\mu$. In the present
calculation we consider a $\mu$ dependent gap resulting from the
solution of the gap equation \cite{alf1}. The resulting QM EOS reads:
\bq
\Omega_{CFL}&=&{6\over \pi^2}\int_0^\nu k^2(k-\mu)\,{\mathrm d}k
\nonumber\\
&+&
{3\over \pi^2}\int_0^\nu k^2(\sqrt{k^2+m_s^2}-\mu)\,{\mathrm d}k-
{3 \Delta^2 \mu^2\over \pi^2}
\eq
with
$
\nu=2\mu-\sqrt{\mu^2+{m_s^2\over 3}}\, ,
$
and the quark density $\rho$ is calculated numerically
by deriving the thermodynamic potential respect to $\mu$. 
Pressure and energy density read:
\bq
P&=&-\Omega_{CFL}(\mu)-B-\Omega^{e}(\mu_e)\\
E/V&=&\Omega_{CFL}(\mu)+\mu\rho+B+\Omega^{e}(\mu_e)+\mu_e\rho_e \, .
\eq

\begin{figure}
\label{masseraggi}
\parbox{6cm}{ \scalebox{0.45}{
\includegraphics*[60,410][550,730]{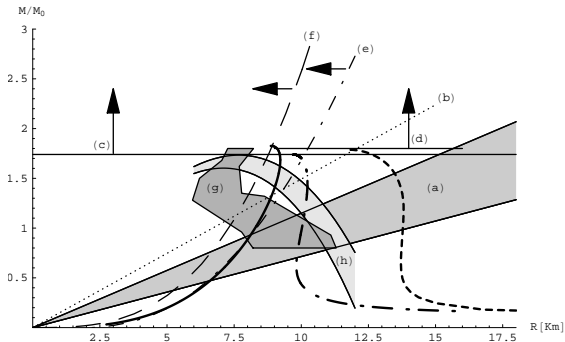}} }

\parbox{9cm}{
\caption{Mass-radius plane with observational limits and representative
theoretical curves: thick solid line indicates CFL quark stars, 
thick dot-dashed line
CFL hybrid stars, thick-dashed line hadronic stars (see text). 
Observational limits from: (a)\,Sanwal et al. 2002 \cite{sanwal},
(b)\, Cottam et al. 2002 \cite{cottam}, (c)\, Quaintrell et al. 2003
\cite{quaint}, (d)\, Heinke et al. 2003 \cite{heinke}, (e),(g)\, Dey
et al. 1998 \cite{dey}, (f)\, Li et al. 1999 \cite{li}, (h)\, Burwitz
et al. 2002 \cite{burwitz}.  }}
\end{figure}

In Fig.1 we have collected most of the analysis of data from X-ray
satellites, concerning masses and radii of compact stellar objects
\cite{sanwal,cottam,quaint,heinke,dey,li,burwitz}. 
Observing Fig.1, we notice that the constraints coming from a few data
sets (labeled ``e'', ``f''\footnote{A very recent reanalysis of the
data of the pulsar SAX J1808.4-3658, discussed in Ref. \cite{li}, seems
to indicate slightly larger radii, of the order of 9-10 km for a star
having a mass of 1.4-1.5 $M_\odot$ \cite{Poutanen}.}  ``g'' and maybe
also constraint ``h''\footnote{In Ref.\cite{zane} an indication for an even more
compact stellar object can be found. Anyway, the so-called thermal radius obtained
in these analysis could be significantly smaller than the total radius of the star})
indicate rather unambiguously the existence of very compact stellar
objects, having a radius smaller than $\sim 10$ km.  At the contrary,
at least in one case (``a'' in the figure), the analysis of the data
suggests the existence of stellar objects having radii of the order of
12 km or larger, if their mass is of the order of 1.4 $M_\odot$.
We recall that it is difficult
from an astrophysical viewpoint to generate compact stellar objects
having a mass smaller than 1 $M_\odot$.  Therefore
the most likely interpretation of constraint ``a'' is that the
corresponding stellar object does not belong to the same class of
objects which have a radius smaller than $\sim 10$ km. Concerning
constraint ``b'', it can be
satisfied both with a very compact star or with a star having a larger
radius.  The apparent contradiction between the constraints ``e'',
``f'', ``g'' and the constraint ``a'' can be easily accommodated in our
scheme, since it can be the signal of the existence of metastable
purely HS which can collapse into a stable configuration
when deconfined QM forms inside the star.

Finally, constraints (``c''\footnote{The result of Ref.\cite{quaint}
is $M/M_\odot = 1.88\pm 0.13$. In Fig.1 only the lower limit is
displayed.}  and ``d''\footnote{If the observed X-ray emission is due
to continuing accretion, a smaller mass is allowed, $M/M_\odot=1.4$.}
) do not provide stringent limits on the radius of the star, but they
put strong constraints on the lower value of its mass.  It is in
general not easy to obtain stellar configurations having both large
masses and very small radii. As we will see, the existence of an
energy gap associated with the diquark condensate helps in
circumventing this difficulty, since the effect of the gap is to
increase the maximum mass of stars having a huge content of
pure QM.

\begin{figure}
\label{gap}
\parbox{6cm}{ \scalebox{0.4}{
\includegraphics*[60,430][560,730]{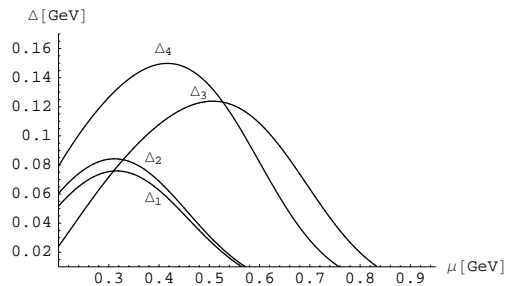}} }

\parbox{9cm}{
\caption{Gap as function of the chemical potential, for four different
parameter sets.}}
\end{figure}

In Fig.1 we show a few theoretical M-R relations which correspond to
the scenario we are proposing. More precisely, we show a 
thick-dashed line corresponding to HSs (GM1), a thick dot-dashed line 
corresponding to HySs
(GM1, $B^{1/4}=170$ MeV, $\Delta_2$) and a thick solid line
corresponding to QSs ($B^{1/4}=170$ MeV, $\Delta_4$).  
Similar shapes can be obtained using the EOS of Ref.\cite{ms96}.
Both the HyS and the QS lines can
satisfy essentially all the constraints derived from observations.
The shapes of the gaps $\Delta_i$ are shown in Fig.~2.
In conclusion, in our scheme most of the
compact stars are either HySs or QSs having a mass in the range
$1.2-1.8 M_\odot$ and a radius $\sim 8.5-10$ km.  
Stars having a significantly larger radius (as the one suggested by constraint ``a'')
correspond in our scheme to metastable HSs which can
exist if their mass is not too large, as we show in the following. 
\begin{table}[b]
\begin{center}
\tabcolsep=0.8\tabcolsep 
\begin{tabular}{ccccccc}
\hline
\hline
Hadronic         &
$B^{1/4}$        & 
$\Delta E$  & 
$\Delta E$  &  
$\Delta E$  & 
$\Delta E$  &
$\Delta E$  \\
Model        &
[MeV]        &
$\Delta=0$   &  
$\Delta_1$   &   
$\Delta_2$   &  
$\Delta_3$   &
$\Delta_4$   \\        

\hline
GM3 & $160$  &  $95$ &  $172^\p$  & $178^\p$  & $204^\p$  & $327^\p$ \\
GM3 & $170$  &  $40$ &  $83$      &  $89$     & $133$     & $236^\p$  \\
GM3 & $180$  &  $10$ &  $29$      &  $31$     & $79$      & --  \\
GM1 & $160$  & $101$ &  $178^\p$  &  $184^\p$ & $210^\p$  & $333^\p$   \\
GM1 & $170$  &  $42$ &  $89$      &  $95$     &  $138$    & $242^\p$  \\
GM1 & $180$  &  $6$  &  $28$      &  $31$     & BH        & --    \\

\hline \hline
\end{tabular}
\caption{Energy released $\Delta E$ (measured in foe=10$^{51}$ erg) in
the conversion from a 1.4 $M_\odot$
hadronic star into the hybrid or quark star having the same baryonic mass (labeled with a $\p$), for
various sets of model parameters.
BH indicates that the hadronic star collapses to a Black Hole.
A dash (--) indicates situations in which the Gibbs
construction does not provide a mechanically stable EOS.
}
\end{center}
\label{tsn}
\end{table}

Let us now discuss $\Delta E$, the energy released in the conversion from HS
to HyS or QS. 
$\Delta E$ is the difference between the gravitational mass of the
HS and that of the final HyS or QS having the same baryonic mass.
As mentioned in the introduction, a possibility is that deconfinement
takes place a few seconds after the bounce, when the PNS has
deleptonized and its temperature has
dropped down \cite{Benvenuto,Pons}. In particular, for stars having a small mass
the formation of QM takes place only at $T\lesssim$ few MeV.
Notice that for a star having a mass of order 1.4$M_\odot$ and using
the relativistic EOSs discussed in this paper, hyperons
are present in the initial configuration, since the typical mass at which hyperons
start forming is $\sim 1 M_\odot$. 
The energy released during quark deconfinement powers a new neutrino
flux which can be useful to make the supernova explode.  
$\Delta E$ is shown in Table I, for a PNS having
a mass of 1.4 $M_\odot$. As it can be seen, $\Delta E$ can be as large as
10$^{53}$ erg, if the final configuration corresponds to an HyS and three times
as large if a QS is obtained. The effect of the gap is to increase the
energy released and to allow QS configurations in cases where an HyS would be
obtained in the absence of quark pairing. 
Let us now remark that
the deconfinement transition can be delayed if a non-vanishing surface tension at the
interface between HM and QM exists and if the mass of the HS is not too large.
This possibility was not discussed in Ref.\cite{Pons} and it is the main ingredient
of our model.
To compute the time needed to form QM we use the technique of quantum tunneling
nucleation.  We can assume that the temperature has no effect in our
scheme because, as discussed above, when QM forms the temperature is so
low that only quantum tunneling is a practicable mechanism.  

In Ref.\cite{noiapj} it was proposed that the central density of a pure
HS (containing hyperons) can increase, due to spin down or mass accretion, until its value
approaches the deconfinement critical density. At this point a
spherical virtual drop of QM can form. The potential energy for
fluctuations of the drop radius $R$ has the form \cite{lif}:
\be
U(R)={4 \over 3} \pi R^3 n_q (\mu_q-\mu_h)+4 \pi\sigma R^2 + 8
\pi\gamma R 
\ee 
where $n_q$ is the quark baryon density, $\mu_h$ and $\mu_q$ are the
hadronic and quark chemical potentials, all computed at a fixed
pressure $P$, and $\sigma$ is the surface tension for the interface
separating quarks from hadrons. Finally, the term containing $\gamma$
is the so called curvature energy.  For $\sigma$ we use standard
values from 10 to 40 MeV/fm$^2$ and we assume that it takes into
account, in a effective way, also the curvature energy. The value of
$\sigma$ was estimated in Ref.\cite{jaffe} to be $\sim 10$
MeV/fm$^2$. Values for $\sigma$ larger than $\sim 30$ MeV/fm$^2$ are
probably not useful at the light of the results of
Refs.\cite{voskresensky,alf2}.

\begin{table}[b]
\begin{center}
\tabcolsep=0.1\tabcolsep 
\begin{tabular}{ccccccccc}
\hline
\hline
Hadronic         &
$B^{1/4}$        & 
$\sigma$         & 
$M_{cr}/M_\odot$ & 
$\Delta E$  & 
$\Delta E$  &  
$\Delta E$  & 
$\Delta E$  &
$\Delta E$  \\
Model        &
[MeV]        &
[MeV/fm$^2$] & 
             & 
$\Delta=0$   &  
$\Delta_1$   &   
$\Delta_2$   &  
$\Delta_3$   &
$\Delta_4$   \\

\hline
GM3 & $170$  & $10$ & $1.12$  & $18$  &  $52$   &  $57$   & $86$ & $178^\p$  \\
GM3 & $170$  & $20$ & $1.25$  & $30$  &  $66$   &  $72$   & $106$ & $205^\p$   \\
GM3 & $170$  & $30$ & $1.33$  & $34$  &  $75$   &  $81$   & $120$ & $221^\p$   \\
GM3 & $170$  & $40$ & $1.39$  & $38$  &  $82$   &  $88$   & $131$ & $234^\p$   \\
\hline
GM3 & $180$  & $10$ & $1.47$  & BH  &  $35$   &  $38$   & BH & --  \\
GM3 & $180$  & $20$ & $1.50$  & BH  &  $38$   &  $40$   & BH & --  \\
GM3 & $180$  & $30$ & $1.52$  & BH  &  $40$   &  $42$   & BH & --   \\
\hline
GM1 & $170$  & $10$ & $1.16$  & $18$ &  $58$  &  $64$   &  $94$ & $189^\p$  \\
GM1 & $170$  & $20$ & $1.30$  & $30$ &  $75$  &  $81$   &  $119$ & $219^\p$ \\
GM1 & $170$  & $30$ & $1.41$  & $43$ &  $90$  &  $96$   &  $141$ & $244^\p$  \\
GM1 & $170$  & $40$ & $1.51$  & BH   &  $105$ &  $111$  &  $163$ & $267^\p$   \\
\hline
GM1 & $180$  & $10$ & $1.56$  & BH  &  $52$     &  $54$   & BH & --    \\
GM1 & $180$  & $20$ & $1.61$  & BH  &  $65$     &  $65$   & BH & --    \\
GM1 & $180$  & $30$ & $1.65$  & BH  &   BH      &   BH    & BH & --     \\

\hline \hline
\end{tabular}
\caption{Energy released $\Delta E$ in
the conversion to hybrid or quark star, for
various sets of model parameters, assuming the hadronic star mean
life-time $\tau=1$ yr (see text). $M_{cr}$ is the gravitational mass
of the hadronic star at which the transition takes place, for fixed
values of the surface tension $\sigma$ and of the mean life-time
$\tau$. Notations as in Tab.~1}
\end{center}
\label{tgrb}
\end{table}

The calculation
proceeds in the usual way: after the computation (in WKB approximation)
of the ground state energy $E_0$ and of the oscillation frequency
$\nu_0$ of the virtual QM drop in the potential well $U(R)$, it is
possible to calculate in a relativistic frame the probability of
tunneling as \cite{iida} 
\be p_0=\exp [-2{A(E_0)\over \hbar}] \ee
where 
\be 
A(E) =\int_{R_-}^{R_+} dR \sqrt{[2 M(R)+E-U(R)][U(R)-E]}  
\ee 
Here $ M(R)=4\pi\rho_h\left(1-{n_q\over n_h}\right)^2 R^3 $,
$\rho_h$ is the hadronic energy density and $n_h$, $n_q$ are the
baryonic densities at a same and given pressure in the hadronic and
quark phase, respectively. Finally, $R_\pm$ are the classical turning points. 
The nucleation time is then equal to 
$\tau = (\nu_0 p_0 N_c)^{-1}$, 
where $N_c$ is the number of centers of droplet formation in the
star, and it is of the order of $10^{48}$ \cite{iida}. 
$\tau$
can be extremely long if the mass of the metastable
star is small enough but, via mass accretion, it can be
reduced from values of the order of the age of the universe down to a
value of the order of days or years. We can therefore determine the
critical mass $M_{cr}$ of the metastable HS for which the nucleation
time corresponds to a fixed small value (1 year in Tab.~1).

In Table II we show the value of $M_{cr}$ for various sets of model
parameters.  In the conversion process from a metastable HS into an
HyS or a QS a huge amount of energy $\Delta E$ is released. We
see in Table II that the formation of a CFL phase allows to obtain
values for $\Delta E$ which are one order of magnitude larger
than the corresponding $\Delta E$ of the unpaired QM case
($\Delta=0$).  Moreover, we can observe that $\Delta E$ depends both on
magnitude and position of the gap.

In the model we are presenting, the GRB is due to the cooling of the
justly formed HyS or QS via neutrino - antineutrino emission.  The
subsequent neutrino-antineutrino annihilation generates the GRB.  In
our scenario the duration of the prompt emission of the GRB is
therefore regulated by two mechanisms: 1) the time needed for the
conversion of the HS into a HyS or QS, once a critical-size droplet is
formed and 2) the cooling time of the justly formed HyS or QS.
Concerning the time needed for the conversion into QM of at least a
fraction of the star, the seminal work by \cite{Oli87} has been
reconsidered by \cite{HB88}, where it has been shown
that the stellar conversion is a very fast process, having a duration
much shorter than 1s. On the other hand, the neutrino trapping time,
which provides the cooling time of a compact object, is of the order
of a few ten seconds \cite{ignazio}, and it gives the typical duration
of the GRB in our model.

In conclusion, comparing the theoretical mass-radius curves with recent
observational data, we find that color superconductivity is a crucial
ingredient in order to satisfy all the constraints coming from
observations. The difficult problem posed by astrophysical
data indicating the existence of stars which are both very
compact and rather massive can be solved either with hybrid or
quark stars.  Concerning hybrid stars, the gap
increases the maximum mass of the stable configuration,
while keeping the corresponding radius $\lesssim$ 10 km. 

The superconducting gap affects also deeply the energy released in the
conversion from hadronic star into hybrid or quark star.  
We assume that the deconfinement transition only takes place when the star has
deleptonized and cooled down, in agreement with the
results of Ref.\cite{Benvenuto,Pons}.
If deconfinement occurs immediately after deleptonization,
the energy released can help the SN to explode. If, at variance, the
transition is delayed, a metastable hadronic star can form.
Its subsequent transition to a stable configuration, containing deconfined quark
matter, can power a GRB via the annihilation of neutrinos
and antineutrinos emitted during the cooling of the newly formed compact star.
The energy released is significantly increased by the 
effect of the chemical-potential dependent superconducting gap and it can reach a value of the order 
of $10^{53}$ erg.
The proposed mechanism could explain
recent observations indicating a possible delay between a Supernova 
and the subsequent Gamma Ray Burst \cite{Amati00,Reeves02,hjorth}.


\end{document}